\documentclass[twocolumn, trackchanges]{aastex61}
\usepackage{natbib}
\usepackage{amsmath}

\submitjournal{ApJL}

\shorttitle{Density fluctuations driven by parametric decay}
\shortauthors{Bowen et al.}

\begin{document}

\title{Density fluctuations in the solar wind driven by Alfv\'en wave parametric decay}

\correspondingauthor{Trevor A.~Bowen}
\email{tbowen@berkeley.edu}

\author{Trevor  A.~Bowen}
\affiliation{Physics Department and Space Sciences Laboratory,
University of California, 7 Gauss Way, Berkeley, CA 94720, USA }%

\author{Samuel Badman}%
\affiliation{Physics Department and Space Sciences Laboratory,
University of California, 7 Gauss Way, Berkeley, CA 94720, USA }%

\author{Petr Hellinger}%
\affiliation{Astronomical Institute, CAS,
Bocni II/1401, CZ-14100 Prague, Czech Republic}%

\author{Stuart D.~Bale}%
\affiliation{Physics Department and Space Sciences Laboratory,
University of California, 7 Gauss Way, Berkeley, CA 94720, USA }%

\newcommand{\bvec}[1]{{\ensuremath{\bf{#1}}}}
\newcommand{\e}[1]{\ensuremath{\times 10^{#1}}}
\def \wind {{\em Wind }}
\def\dens[#1]{10$^{#1}$\hskip 1.5pt{cm$^{-3}$}}

\def\ex {\ensuremath {\hat{{\bvec{e}}}_1}}
\def\ey {\ensuremath {\hat{{\bvec{e}}}_2}}
\def\ez {\ensuremath {\hat{{\bvec{e}}}_3}}
\def\epar {\ensuremath {\hat{{\bvec{e}_\parallel}}}}

\def\cordnb {\ensuremath {C(\delta n, \delta B_\parallel)}}
\begin{abstract}
Measurements and simulations of inertial compressive turbulence in the solar wind are characterized by anti-correlated magnetic fluctuations parallel to the mean field and density structures. This signature has been interpreted as observational evidence for non-propagating pressure balanced structures (PBS), kinetic ion acoustic waves, as well as the MHD slow-mode. Given the high damping rates of parallel propagating compressive fluctuations, their ubiquity in satellite observations is surprising, and suggestive of a local driving process. One possible candidate for the generation of compressive fluctuations in the solar wind is Alfv\'en wave parametric instability. Here we test the parametric decay process as a source of compressive waves in the solar wind by comparing the collisionless damping rates of compressive fluctuations with growth rates of the parametric decay instability daughter waves. Our results suggest that generation of compressive waves through parametric decay is overdamped at 1 AU, but that the presence of slow-mode like density fluctuations is correlated with the parametric decay of Alfv\'en waves.
\end{abstract}
\section{introduction}

Observations of the solar wind contain a characteristic anti-correlation between density fluctuations and magnetic perturbations parallel to the mean field.  \citet{TuMarsch1994} interpreted observed correlations on hourly timescales between magnetic pressure, thermal pressure, and density fluctuations from {\em{Helios}} as non-propagating pressure balance structures (PBS).  Using spacecraft potential measurements as a proxy for density fluctuations, \citet{KelloggHorbury2005} reported the same anti-correlation in {\em{Cluster}} data, noting that PBS are consistent with the perpendicular limit of the ion-acoustic wave. \citet{Yaoetal2011}, using wavelet coherence analysis, demonstrated this anti-correlation through the range of inertial scales; subsequent work interpreted these observations as PBS driven by either slow-modes or mirror modes arising from temperature anisotropies \citep{Yaoetal2013a,Yaoetal2013b}. Through statistical analysis of a decade of \wind observations, \cite{Howesetal2012} attributed the compressible component of magnetic fluctuations entirely to the kinetic slow-mode. However, comparisons between a decade of \wind observations and numerical predictions made in both kinetic and MHD simulations by \citet{Verscharenetal2017} suggest that the MHD slow-mode best explains the compressible component of the solar wind.

Strong collisionless Landau damping of kinetic slow-mode waves propagating parallel to the magnetic field is expected for solar wind parameters at 1 AU \citep{Barnes1966}. For this reason, the generation, presence, and damping of kinetic slow-mode fluctuations in the solar wind remains an open question in space physics. \cite{Kleinetal2012} argued that the distribution of slow-mode waves in the solar wind is consistent with critically balanced turbulence, in which non-linear interactions between Alfv\'enic and slow-mode fluctuations cascade weakly damped perpendicular compressible fluctuations at large scales to smaller parallel scales. The damping of these waves could provide a source of heating in the solar wind \citep{NaritaMarsch2015}.

In this letter, we explore the parametric decay instability (PDI) as a process for generating compressive fluctuations in the solar wind. In MHD, the PDI is recognized as the instability of circularly polarized Alfv\'en eigenmodes at high fluctuation amplitudes \citep{Derby1978,Goldstein1978}. For low $\beta$ plasmas, the large amplitude Alfv\'enic fluctuation (mother wave) couples to two daughter waves: a parallel propagating compressive fluctuation, and a backwards propagating Alfv\'en wave. A class of parametric instabilities exist outside the low $\beta$ PDI decay: e.g. at $\beta \eqslantgtr 1$ the instability is dominated by the growth of forward propagating Alfv\'{e}nic daughter waves at twice the mother wave frequency \citep{JayantiHollweg1993,Hollweg1994}.

Simulations of high amplitude circularly polarized Alfv\'en waves have verified the PDI process in the MHD regime, and have suggested that PDI daughter waves can seed turbulent cascades \cite{DelZannaetal2001,TeneraniVelli2013, Malaraetal2001}. Simulations have extended the PDI to large amplitude Alfv\'en waves with arc and linear polarizations, non-monochromatic wave distributions, and oblique propagation \cite{DelZanna2001,Matteinietal2010, DelZannaetal2015}. Furthermore, kinetic simulations recover PDI in multi-species models, including effects such as, the preferential heating of $\alpha$ particles, proton core heating, and beam formation \cite{Aranedaetal2008,Aranedaetal2009,KauffmanAraneda2008}. Numerical simulations of PDI in a turbulent solar wind demonstrate the generation of slow-mode fluctuations from the decay of Alfv\'enic fluctuations in a relatively robust theoretical scenario \cite{Shietal2017}.

%
%We implement a decade of measurements from the \wind mission to investigate whether compressive slow-mode fluctuations in the solar wind could be generated by the parametric decay of Alfv\'enic fluctuations. Signatures of compressive parallel propagating slow-mode waves with large damping rates are readily observed in this data. We demonstrate that the distribution of {\em{in-situ}} observations is bounded by high parametric decay growth rates. Furthermore we find that intervals of solar wind with large PDI growth rates of correlate well with density fluctuations, suggesting that the PDI may drive the growth of density fluctuations in the propagating solar wind.
%
\section{Data}
We use 10 years of solar wind measurements at 1AU from the NASA \wind mission ranging from 1996 January 1 through 2005 December 31. Data from the Magnetic Field Investigation (MFI) \cite{LeppingMFI}, Solar Wind Experiment (SWE) \cite{OgilvieSWE}, and Three Dimensional Plasma (3DP) experiment \cite{Lin3DP} are separated into non-overlapping 300 second intervals. Intervals are excluded when {\em{Wind's}} geocentric distance is less than $35 R_E$, the average solar wind speed is $<$ 250 km/s, or if an interval is missing $>15\%$ of coverage from any instrument. Data gaps are interpolated when $ <15\%$ of measurements are missing.  Additionally, intervals are excluded if there is significant discrepancy between 3DP and SWE measurements of mean proton density such that  $|(n_{SWE}-n_{3DP})|/n_{SWE} > 0.1$. The resulting data set consists of 533222 individual 300 second intervals. Techniques from \cite{Pulupaetal2014} were used to obtain electron densities and temperatures for 282286 of these intervals.

The 3 s cadence 3DP ``on board" proton moment measurements are interpolated to the MFI time base. For each interval, measurements of density, velocity, and magnetic field (represented as $n$, $\bvec{v}$, and $\bvec{B}$) are separated into mean and fluctuation quantities using time averaged quantities, denoted as $\langle...\rangle.$ For example, the mean magnetic field, $\bvec{B_0},$ is determined through $\langle\bvec{B}\rangle=\bvec{B_0}$ while the fluctuation field is determined through $\bvec{B}=\bvec{B_0}+\bvec{\delta B}$. Vector fluctuations are rotated into a field aligned coordinate (FAC) system, $\delta\bvec{B}_{FAC}=(\delta B_{\perp1}, \delta B_{\perp2}, \delta B_{\parallel}),$ with the parallel fluctuations, $\delta B_{\parallel}$, defined along the mean field direction $\bvec{B_0}/|B_0|$. 

Each interval is characterized through root-mean-square (RMS) quantities normalized to the mean value $$\bar{\delta n}=\sqrt{\langle\delta n^2}\rangle/n_0.$$ Vector fluctuations characterizing each interval separated into parallel and perpendicular RMS values of the form $$\bar{\delta B_\perp}= \frac{\sqrt{\langle\delta B^{2}_\perp\rangle}}{B_0}=\frac{\sqrt{\langle\delta B_{\perp 1}^2+\delta B_{\perp 2}^2\rangle}}{B_0}$$ and $$\bar{\delta B_\parallel}=\frac{\sqrt{\langle\delta B_\parallel^2\rangle}}{B_0}.$$

The FAC fluctuation quantities are used in deriving Els{\"a}sser variables, $$\bvec{z}^{\pm}=\bvec{\delta v}_{FAC} \pm\bvec{\delta b}_{FAC};$$
normalized cross helicity, $$H_c=\frac{\langle\delta\bvec{b}_{FAC}\cdot \delta\bvec{v}_{FAC}\rangle}{\langle{\delta b}^2\rangle+\langle{\delta v}^2\rangle};$$
 and the zero lag cross correlation between parallel field fluctuations and density, $$C(\delta n, \delta B_\parallel)=\frac{\sum_{i=0}^{N-1} \delta n_i\delta B_{\parallel i}}{\sqrt{\sum_{i=0}^{N-1}{\delta n_i}^2}\sqrt{\sum_{i=0}^{N-1}{\delta B_{\parallel i}}^2}}.$$ 
 The magnetic field has been normalized as $\bvec{\delta b}=\bvec{\delta B}/\sqrt{\mu_ 0n_p m_p}$.

\section{Model Propagation Direction}

%%
%%		Figure 1 - Hc and XCC vs theta_sm
%%

%\begin{figure*}
%\plotone{figs/parametric_ms_fig1.eps}
\begin{figure*}
\includegraphics[width=18cm]{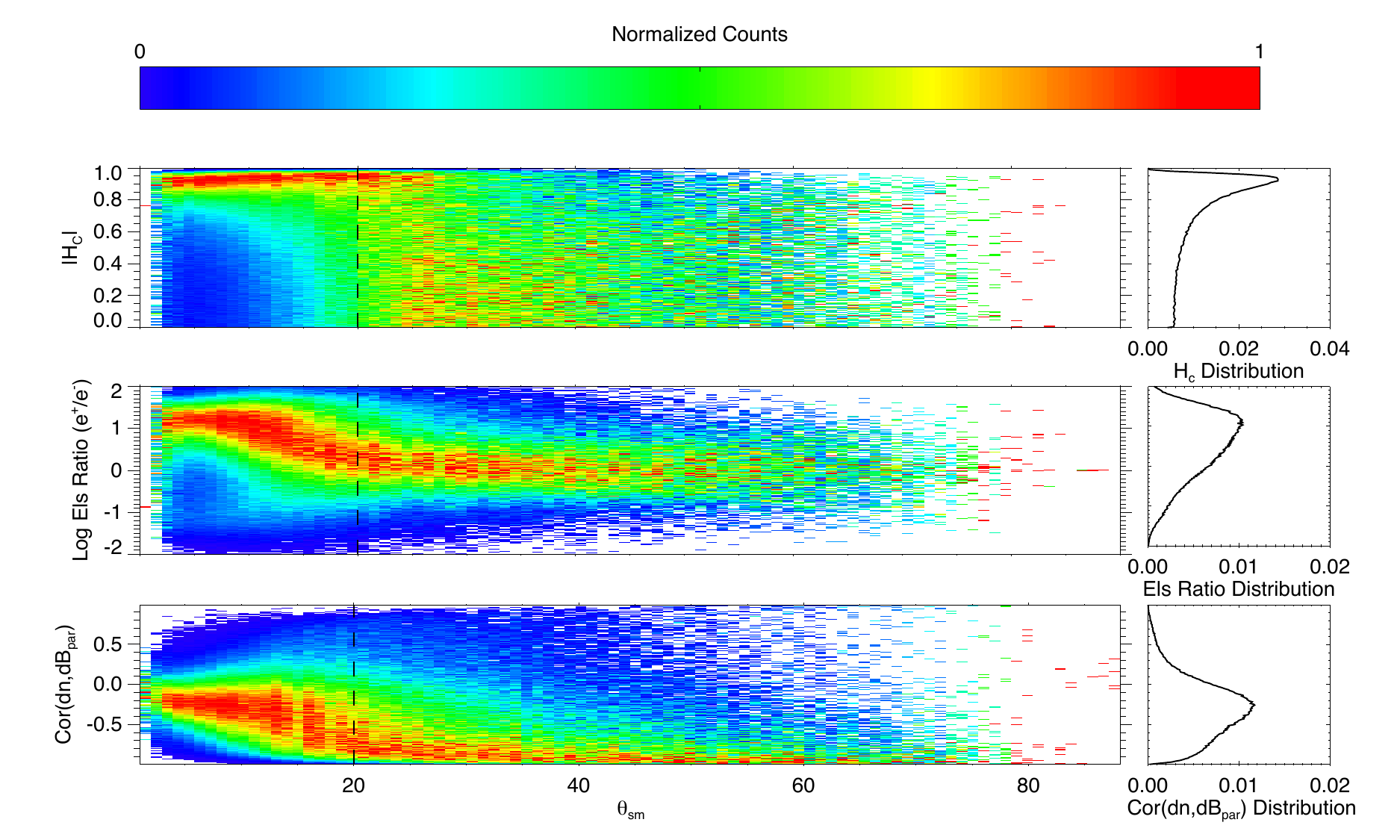}
\caption{(Top) Joint distribution of normalized cross helicity, $H_c$ and slow-mode propagation angle $\theta_{sm}$. Data are column normalized within each angle bin (1$\degr$ resolution). The distribution of $H_c$ observations shown on the right. (Middle) Column normalized distribution of the Els\"asser ratio and $\theta_{sm}$. The distribution of Els\"asser ratios is plotted to the right. (Bottom) Joint distribution of $\cordnb$, and the slow-mode propagation angle, $\theta_{sm}$. The data are column normalized and the distribution of $\cordnb$ observations plotted to the right. A dashed line identifies a transition observed in all three panels at $\theta_{sm} \sim 20^{\circ}.$ \label{fig:one}}
\label{fig:fig1}
\end{figure*}

%
%Calculating the propagation directions for shear MHD Alfv\'en waves is typically implemented with a minimum variance analysis  \citep{Yaoetal2013a}. 
%%For a shear MHD Alfv\'en wave with $\bvec{B_0}=(0,0,B_0)$ and propagation vector in the \ex-\ez plane such that $\bvec{k}=(k_\perp,0,k_\parallel)$ it can be shown that $\bvec{\delta B}=(0, \delta B_\perp,0)$, $\bvec{\delta V}=(0, \delta V_\perp,0)$, and $\bvec{k}=(0,0,k_\parallel)$. 
%The minimum variance direction is acquired through diagonalizing the covariance matrix of $\bvec{\delta B}$. As the shear Alfv\'en wave consists of only transverse fluctuations, the minimum eigenvalue direction yields an estimate of the shear Alfv\'en propagation direction.
%%
For slow-mode waves with propagation vector, \bvec{k}, such that $\bvec{k}=(k_{\perp1},0,k_\parallel)$ it follows that $\bvec{\delta v}=(\delta v_{\perp1}, 0, \delta v_\parallel)$. From the MHD induction equation {$\omega\bvec{\delta B}=(\bvec{k}\cdot \bvec{\delta v})\bvec{B_0} -(\bvec{k}\cdot{\bvec{B_0}})\bvec{\delta v},$} it is derived that $$\text{tan}\theta_{sm}=\left|\frac{k_\perp}{k_\parallel}\right|=\left|\frac{\delta B_\parallel}{\delta B_\perp}\right|,$$ where $\theta_{sm}$ gives the slow-mode propagation angle relative to the mean magnetic field. The limit of small $\delta B_\parallel$, corresponds to quasi-parallel propagation of the slow-mod; for exact parallel propagation, $\delta B_\parallel=0$, the compressive component of the slow-mode becomes purely acoustic.

Figure 1 (top) shows joint probability distributions of $|H_c|$ and $\theta_{sm}$ normalized to the maximum counts in each column. For a shear Alfv\'en wave $H_c =\pm 1$; our observations show that, for small propagation angles, the majority of intervals have a cross-helicity similar to shear Alfv\'en waves. 

Figure 1 (middle) shows the column normalized joint distribution of $\theta_{sm}$ and the ratio of Els\"asser energies: $e^+/e^-$, where $e^{\pm}=\langle (z^{\pm})^2\rangle.$ Intervals consisting of pure outward propagating Alfv\'en waves have large Els\"asser ratios: i.e. intervals with $z^+\gg z^-$ correspond to $e^+ \gg e^{-}$. Intervals containing balanced compositions of $z^{-}$ and $z^{+}$ fluctuations have Els\"asser ratios $\sim$ 1. We find that intervals with small propagation angles are dominated by $z^+$ fluctuations. Intervals with larger $\theta_{sm}$ show balanced  $z^+$ and $z^-$ energies, suggesting that these intervals are not easily represented as single wave fluctuations, but must be modeled by either counter-propagating Alf\'enic fluctuations or a quasi-linear superposition of wave-modes \citep{Kleinetal2012}.

Figure 1 (bottom) shows the the column normalized joint distribution of $\cordnb$ and $\theta_{sm}.$ The transition to highly anti-correlated density $\delta n$ and parallel magnetic fluctuations $\delta B_{||}$ occurs at approximately the same propagation angle, $\theta_{sm} \sim20^\circ,$ as cross helicity decreases and the Els\"asser energies balance. 

Two regimes are apparent: the first consists of intervals characterized by large cross helicity and `outward' propagating energies. The second regime consists of highly compressible intervals with mixed cross helicity, balanced Els\"asser energies, and strong density-field correlations consistent with slow-mode fluctuations. \citet{Howesetal2012} restricted analysis to intervals with $\delta n > {0.5}$ cm$^{-3}$, arguing that lower density intervals may be subject to instrumental noise; the uniform and simultaneous transitions in $H_c$, $e^+/e^-$, and $\cordnb$ with $\theta_{sm}$ without restricting density amplitudes suggests that our analysis is insensitive to noise due to low amplitude densities.

\section{Damping}

\begin{figure}[h]
%\plotone{figs/parametric_ms_fig2.eps}
\includegraphics[width=9cm]{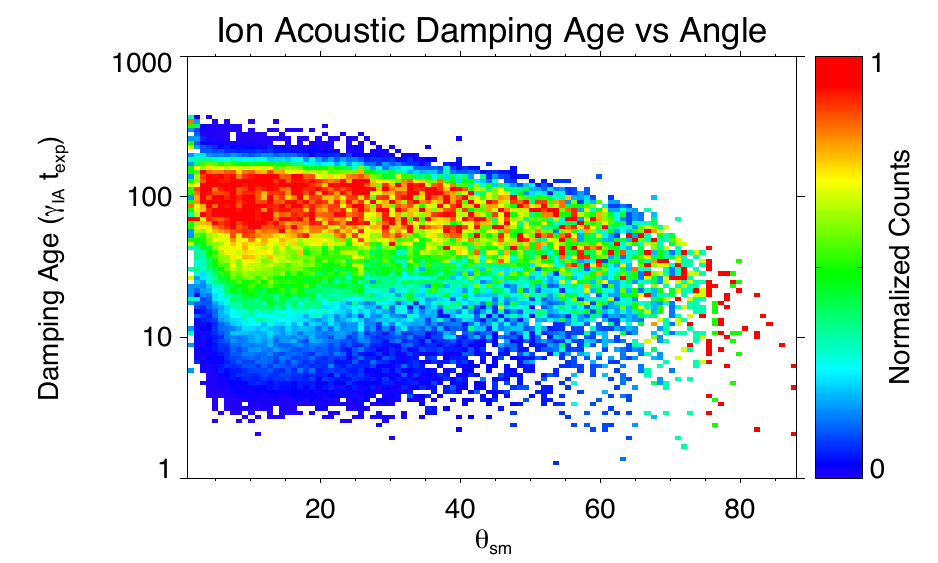}
\caption{Joint distribution of expansion normalized ion-acoustic damping rates, $\gamma_{IA}t_{exp}$, ($t_{exp}=1\text{AU}/V_{sw}$) and slow-mode propagation angle $\theta_{sm}$. Data are restricted to intervals with fitted electron temperatures. The distribution is column normalized in each angle bin (1$\degr$ resolution). Damping ages, $\gamma_{IA}t_{exp}$, are uniformly greater than unity.}
\label{fig:two}
\end{figure}

Historically, the presence of kinetic slow-mode fluctuations in the solar wind has been questioned due to the heavy collisionless damping expected for solar wind parameters at 1 AU. \citep{Barnes1966}.  However, Figure 1 (bottom) reveals the presence of compressive fluctuations with quasi-parallel propagation (e.g. $\theta_{sm} \sim 20\degr$). \citet{Verscharenetal2017} previously argued that observations of the compressive solar wind plasma are best represented as oblique, weakly damped, MHD slow-mode waves which propagate more freely through the heliosphere; we adopt their expression for ion acoustic (IA) damping rates derived for Maxwellian particle distributions:

\begin{equation}
\gamma_{IA}/\omega_s \simeq -\sqrt{\pi}\frac{c_s^3}{w_{\parallel p}^3}\frac{e^{-{c_s^2/w}^2_{\parallel p}}}{1+3w_{\parallel p}^2/c_s^2},\end{equation}
where the ion acoustic frequency is $\omega_s=k_\parallel c_s$, the parallel thermal proton speed is $w_{\parallel p} =\sqrt{2k_BT_{\parallel p}/m_p},$ the ion-acoustic speed is $$c_s=\sqrt{\frac{3k_BT_{\parallel p} + k_BT_{\parallel e}}{m_p}},$$
 $k_\parallel =k\text{cos}\theta_{sm}$, $m_p$ is the proton mass, and $T_{\parallel j}$ is the parallel temperature for species $j$.
 
Figure 2 shows IA damping rates normalized to the solar wind expansion time $t_{exp}=1\text{AU}/V_{sw}$ as a function of $\theta_{sm}$. Data are restricted to intervals with fits of electron distributions \citep{Pulupaetal2014}. The majority of expansion normalized damping ages are above 10, and uniformly above unity, suggesting that even highly oblique propagating kinetic slow-mode like fluctuations undergo significant damping over 1 AU propagation.

 %%
%%		Figure 3 - damping age vs theta_sm
%%

\begin{figure}[h]
%\plotone{figs/parametric_ms_fig3.eps}
\includegraphics[width=9cm]{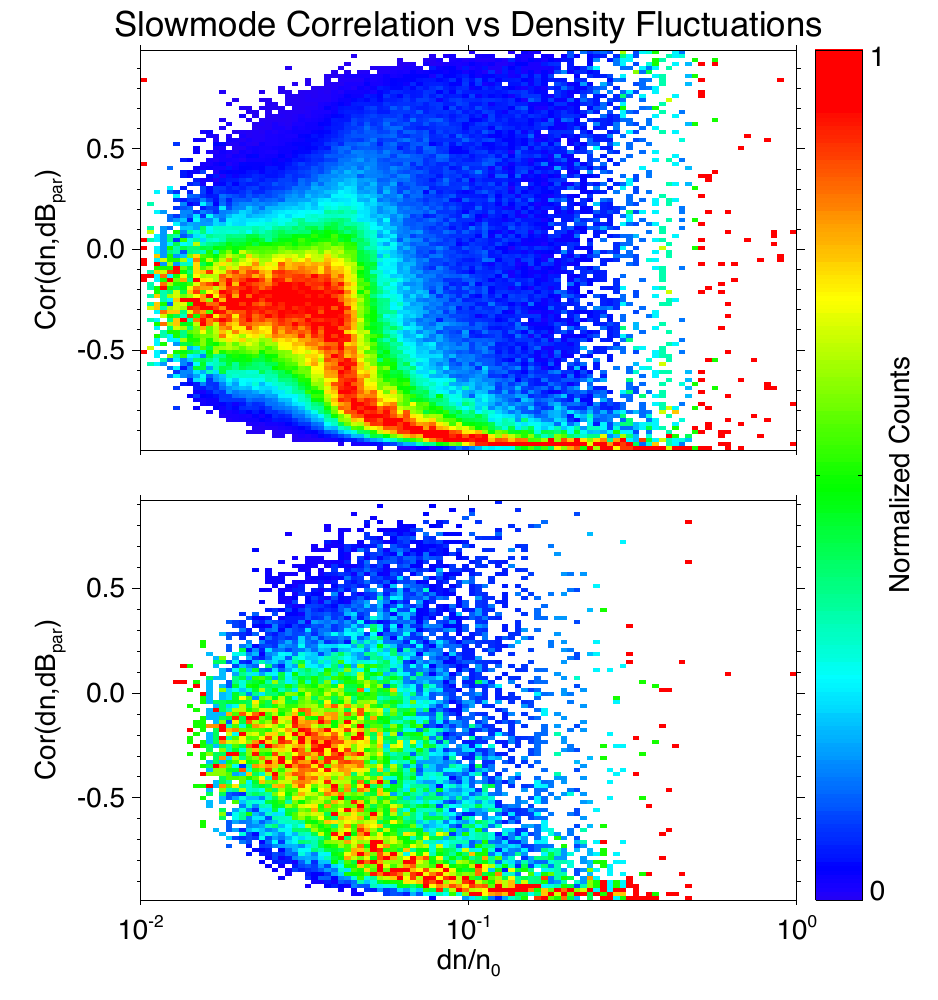}
\caption{(Top) Joint probability distribution of density-parallel magnetic fluctuation cross-correlation, $\cordnb$, and density fluctuation amplitude, $\bar{\delta n}$. (Bottom) Restricting data to intervals with $\beta < 0.5$ and $\theta_{sm} < 20{\degr}$ reveals the presence of fluctuations with slow-mode like density-field correlations propagating quasi-parallel to the mean magnetic field. Data are column normalized to the maximum counts of $\cordnb$ in each $\bar{\delta n}$ bin. \label{fig:three}}
\label{fig:fig3}
\end{figure}

 Figure 3 (top) shows the column normalized joint distribution of $\bar{\delta n}$ and $\cordnb$ while Figure 3 (bottom) shows that the slow-mode correlation emerges even for quasi-parallel propagation ($\theta_{sm}<20{\degr}$). If these fluctuations are associated with a strongly damped kinetic slow-mode, then their presence at 1 AU demands a local driving process. 
 \section{Parametric Decay}

%%
%%		Figure 4 - dB vs beta
%%

\begin{figure*}[h]
%\plotone{figs/parametric_ms_fig4.eps}
\includegraphics[width=18cm]{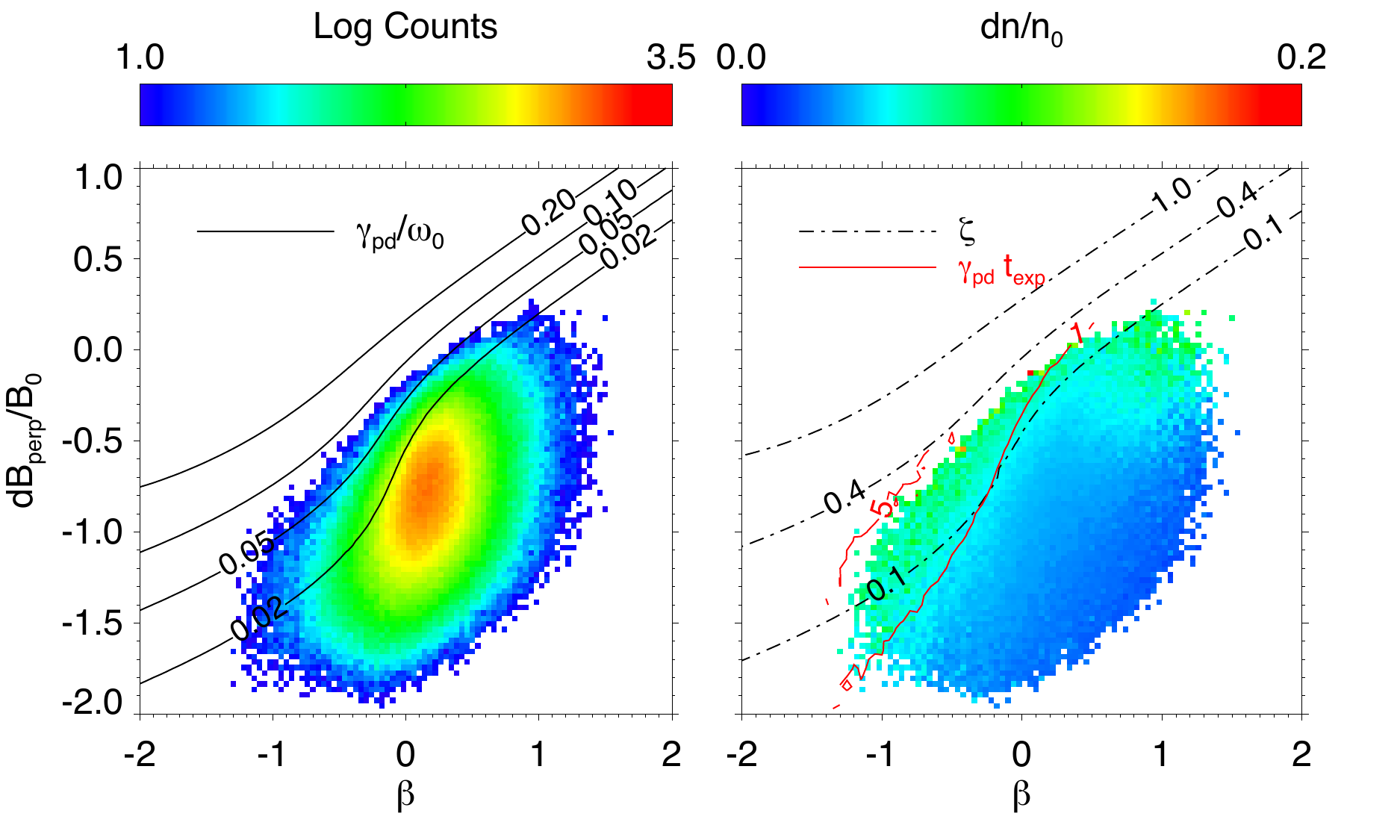}
\caption{(Left) Joint distribution of $\bar{\delta B_\perp}$ and $\beta$. Contours of the  parametric growth rate normalized to the linear wave frequency are shown in solid black. Data are bounded by high PDI growth rates. (Right) Joint  distribution of $\bar{\delta B_\perp}$ and $\beta$ with color scale given by mean values of $\bar{\delta n}.$ Contours of the expansion normalized parametric decay rate, $ \gamma_{pd} t_{exp}$ shown in red, suggest that several iterations of the decay may occur over 1 AU propagation. Contours of $\zeta=\frac{\gamma_{pd}/\omega_0}{\gamma_{IA}/\omega_s}$ are shown in dashed black lines. Though large density fluctuations correspond to the largest values of $\zeta$, generation of density fluctuations through PDI is overdamped.}
\label{fig:fig4}
\end{figure*}

To test whether parametric coupling between wave modes generate compressive fluctuations from the decay of large amplitude Alfv\'enic fluctuations we compare our observations with analytically derived parametric growth rates. Following \cite{Derby1978} and \cite{Goldstein1978}, a dispersion relation for the evolution of a compressive wave coupled to an MHD mother wave is given by:

\begin{multline}
(\omega + k + 2)(\omega + k -2)(\omega - k)(\omega^2 - \beta k^2) \\ -
\bigg(\frac{\delta B_\perp}{B_0}\bigg)^2k^2(\omega^3 + k\omega^2 - 3 \omega + k) =0.
\end{multline}

This $5^{\text{th}}$ order polynomial, with $k$ and $\omega$ corresponding to wave number and frequency of the compressive daughter wave normalized to mother wave quantities, depends only on $\beta$ and $\delta B_\perp/B_0$. There is a small range of $k_L<k<k_U$  for which a single conjugate pair of complex solutions exist. The imaginary part of the solution, i.e.~ the parametric decay rate $\gamma_{pd}$, corresponding to the fastest growing decay product is determined numerically given observed values of $\beta$ and $\bar{\delta B_\perp}.$

Figure 4 (left) shows the joint distribution of $\bar{\delta B_\perp}$ and $\beta$. Contours of numerically determined $\gamma_{pd}/\omega_{0}$, plotted as solid black lines, bound the data at $\gamma_{pd}/\omega_{0} \sim 0.05-0.1$, suggesting that the solar wind at 1 AU is bounded by PDI growth. Figure 4 (right) shows the same distribution of $\bar{\delta B_\perp}$ and $\beta$ with the average value of $\bar{\delta n}$ projected onto the plane. The median $\bar{\delta n}$ is 0.034, ranging between [0.01, 1.10]. Our results show that intervals with the largest density fluctuations correspond to intervals of high parametric growth rates.

Two sets of contours are overlaid in Figure 4 (right). The first set, $\gamma_{pd} t_{exp}$ plotted in red, correspond to PDI growth rates normalized to solar wind expansion time. Normalizing by expansion timescale requires estimating $\gamma_{pd},$ with physical units of s$^{-1},$ from contours of $\gamma_{pd}/\omega_0.$ Using $$\omega_0= k_\parallel V_{alf},$$ where $k$ is given by the Taylor hypotheses, ${k=1/(\tau V_{sw})}$, where $\tau$ is the interval duration (300 s), and $k_\parallel=k \text{cos}\theta,$ it follows that, \begin{equation}\gamma_{pd} t_{exp}= \frac{\gamma_{pd}} {\omega_0} \frac{1 \text{AU}}{\tau} \frac{V_{alf}}{V_{sw}^2} \text{cos} \theta. \end{equation} Contours of $\gamma_{pd}/\omega_0$, are normalized using mean values of $V_{sw}$ and $V_{alf}$ in each bin of the joint $\bar{\delta B_\perp}$ and $\beta$ distribution. We choose an oblique propagation angle of $\theta=80^\circ$ in accordance with observations of anisotropy in the solar wind \citep{Horburyetal2008}. At oblique propagation angles $\text{cos}\theta$ varies weakly with $\theta$, implying that our results are robust to specific oblique angle chosen. Previous work by \cite{Matteinietal2010} and \cite{DelZannaetal2001} has shown that the decay of oblique waves is analogous to parallel decay, scaling as $\text{cos}\theta$; furthermore, \cite{Chandran2017} demonstrate that non-linear PDI interactions for oblique waves occur between parallel components, generating an inverse parallel cascade of Alfv\'en waves.

Figure 4 (right) shows that intervals undergoing a single iteration of parametric decay demonstrate enhanced density fluctuations, suggesting that density fluctuations are correlated with growing compressive PDI daughter waves. Though parametric decay rates measured at {\em{Wind}} are fairly small, propagation over 1AU allows for growth of the instability.  Additionally, contours of $\gamma_{pd} t_{exp}$ show that PDI growth could be smaller than our theoretical rates for circularly polarized, parallel propagating waves, and still generate compressive fluctuations at 1AU.

%
%Two sets of contours are overlaid in Figure 4 (right). The first set, $\gamma_{pd} t_{exp}$ plotted in red, correspond to PDI growth rates normalized to the solar wind expansion time. Normalizing by expansion timescale requires estimates for $\gamma_{pd}$ with physical units of s$^{-1}$.  These are constructed from contours of $\gamma_{pd}/\omega_0,$ using $$\omega_0= \text{median}(V_{alf})\cdot\text{median}(k_\parallel).$$ Here $k$ is given by the Taylor hypotheses, ${k=1/(\tau V_{sw})}$, where $\tau$ is the interval duration (300 s), and $k_\parallel=k \text{cos}\theta$.  We consider only median values of intervals with $\beta < 0.5$ where PDI is most relevant and the MHD slow-mode is approximated by IA fluctuations. Additionally, due to known wave-vector anisotropy in the solar wind, we choose an oblique propagation angle of $\theta=80^\circ$ \citep{Horburyetal2008}. Previous work by \cite{Matteinietal2010} and \cite{DelZannaetal2001} has shown that the decay of oblique waves is analogous to parallel decay, scaling as $\text{cos}\theta$; furthermore, \cite{Chandran2017} demonstrate that the non-linear PDI interaction for oblique waves occurs between their parallel components, leading to an inverse parallel cascade of Alfv\'en waves.
%  

Uncertainty in contours of $\gamma_{pd} t_{exp}$ are determined through the fractional error of Equation 3, derived as \begin{multline}\frac{\Delta(\gamma_{pd} t_{exp})}{\gamma_{pd} t_{exp}}=\\
\sqrt{\left(4\frac{\Delta V_{sw}}{<V_{sw}>}\right)^2+\left(\frac{\Delta V_{alf}}{<V_{alf}>}\right)^2 -4\frac{\Delta V_{sw}\Delta V_{alf}}{V_{sw}V_{alf}}}\end{multline}

The ensemble mean values follow as $<V_{sw}>=453$ {km s$^{-1}$} and $<V_{alf}> =60$ {km s$^{-1}$}, with standard deviations of $\Delta V_{sw}=108$ {km s$^{-1}$} and  $\Delta V_{alf}=34$ {km s$^{-1}$} respectively. Measurements of $V_{alf}$ and $V_{sw}$ are correlated with a value of $\rho=0.38$. Evaluating ${\Delta(\gamma_{pd} t_{exp})}/{\gamma_{pd} t_{exp}}$ gives fractional uncertainty on the contours of ~60\%. For a single instance of the parametric decay, $\gamma_{pd} t_{exp} =1$, uncertainty of 60\% does not significantly impact the contour location.

The second set of contours in Figure 4 (right), $\zeta=\frac{\gamma_{pd}/\omega_0}{\gamma_{IA}/\omega_s}$, represent the ratio of PDI growth rates to IA damping rates. High values of $\zeta$ presumably indicate the presence of driven compressive daughter waves. Intervals with low values of $\zeta$ should dissipate compressive fluctuations faster than the PDI can drive them. Contours of $\zeta$ are constructed by normalizing the contours of ${\gamma_{pd}/\omega_0}$ to the observed median dimensionless damping rates, $\text{median}(\gamma_{IA}/\omega_s)$. For intervals with available electron fit data and $\beta < 0.5$ the median value of ${\gamma_{IA}/\omega_s}$ is 0.26, with a standard deviation of $\sim0.05$. This leads to a fractional error to the contours of $\Delta\zeta/\zeta \sim15\%$. Figure 4 (right) shows that though intervals with large density fluctuations occur with greater values of $\zeta$, compressive fluctuations driven by parametric decay are overdamped, i.e. have PDI growth rates less than IA damping rates.

\begin{figure}[h]
%\plotone{figs/parametric_ms_fig5.eps}
\includegraphics[width=9cm]{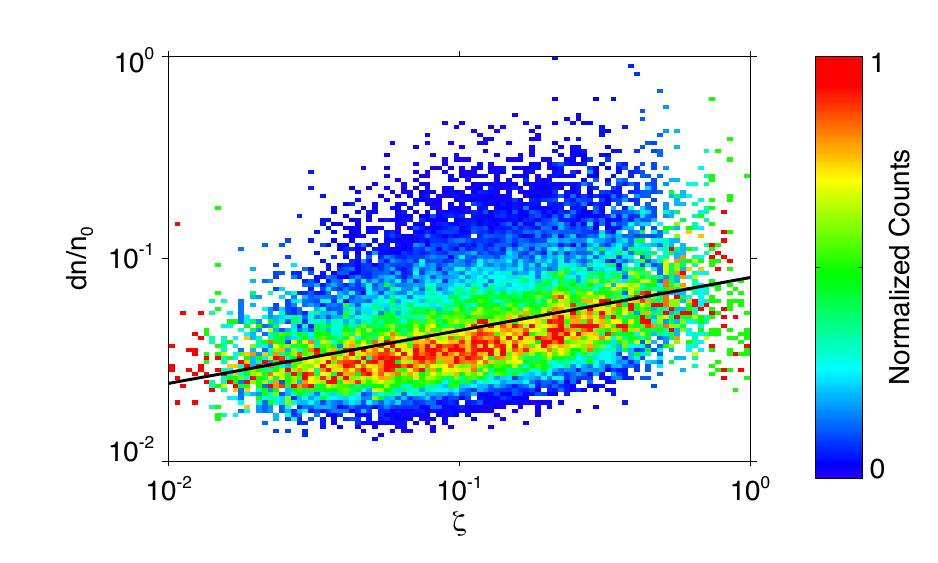}
\caption{Distribution of $\bar{\delta n}$ and $\zeta$. Data are column normalized to each $\zeta$ bin. A least squares fit to the data, shown as a black line, gives a power law of 0.25 and a Pearson correlation of 0.36.}
\label{fig:fig5}
\end{figure}

Figure 5 shows the scaling of $\bar{\delta n}$ with $\zeta$ computed for each interval. Data are restricted to $\beta <0.5$, where PDI generates compressive daughter waves. A positive correlation is observed between the density fluctuations and $\zeta.$ A least squares fit in logarithmic space suggests a power law relationship of $\bar{\delta n}\sim(\zeta)^{0.26}$ with a Pearson cross correlation of $\rho=0.36$; again we note that approximately 99\% of the data are overdamped. When a least square fit is performed between $\bar{ \delta n}$ and the dimensionless parametric growth rate, $\gamma_{pd}/\omega_0$ we recover a power law relation of $\bar{ \delta n}\sim (\gamma_{pd}/\omega_0)^{0.26}$ and a Pearson correlation of $\rho=0.37.$ Suggesting that density fluctuations $\bar{ \delta n}$ are weekly dependent on the IA damping rate. Additionally, the positive correlation between parametric instability growth rate and density fluctuations in the solar wind suggests that compressive fluctuations may be related to PDI. Both of these correlations are stronger than the correlation between amplitudes of $\bar{\delta B_\perp}$ and $\bar{\delta n}$, for which we find $\rho=0.24.$

 \section{Discussion}
 
Our results show that the compressive component of solar wind plasma is not restricted to fluctuations propagating perpendicular to the magnetic field. If these compressive fluctuations correspond to a kinetic slow-mode, their presence likely requires a driving process. Previous work has invoked a scalar turbulent cascade to explain the replenishing of slow-mode like fluctuations \citep{Kleinetal2012, Chen3danis2012}. 

In this letter, we suggest that parametric coupling of large amplitude Alfv\'{e}n waves to compressional modes may exist in the solar wind. We consider parametric decay as a low $\beta$ parallel process; however, the decay of oblique shear Alfv\'{e}n waves generates a spectrum of fluctuations over a range of angles relative to the mean magnetic field \citep{Matteinietal2010}. One complication is that the parametric growth rates adopted from \cite{Derby1978} are relevant for low $\beta$ plasmas; however, we note that \cite{Teneranietal2017} demonstrate that pressure anisotropies allow PDI growth at higher $\beta.$ 

Parametric decay has been verified in analytic and computational studies but it has yet to be observed in the solar wind. Parametric coupling is an attractive physical mechanism for explaining compressive fluctuations in the solar wind for two reasons. First, backwards propagating Alfv\'enic fluctuations, a product of PDI, are required for non-linear interactions leading to critically balanced turbulence \citep{TeneraniVelli2013,Malaraetal2001}. Second, parametric coupling preferentially excites slow-mode fluctuations; which may explain the distribution of fluctuations observed at 1 AU: Alfv\'enic $\sim 90\%$, slow-mode $\sim 10\%$, and very little fast-mode contribution \citep{Kleinetal2012}. 

The results presented in Figure 1, in which slow-mode fluctuations are observed simultaneously with balanced (i.e. equally distributed backwards and forwards propagating Alfv\'enic fluctuations) intervals provide compelling observational evidence for parametric decay. Furthermore, observations from \wind show that high PDI growth rates bound the distribution of data, suggesting that parametric coupling may limit the amplitude of Alfv\'enic fluctuations. The results in Figure 4 show that intervals of solar wind which undergo a single growth of parametric decay have enhanced density fluctuations.

Though the IA damping term given in Equation 1 is greater than PDI growth rates, we note that non-Maxwellian particle distributions in the solar wind may significantly lower the IA damping rates. In particular, we highlight the results of \cite{Verscharenetal2017}, which suggest that slow-mode fluctuations in the solar wind are fluid like with low damping rates. Reducing IA damping by approximately a factor of 5 would provide underdamped growth of compressive fluctuations through PDI.

Additionally, scaling arguments give much higher PDI rates in the inner heliosphere, increasing generation of compressive waves through PDI near the sun \citep{TeneraniVelli2013}. The launch of the NASA Parker Solar Probe mission in July 2018 will reveal whether parametric coupling between fluctuations play an increased role in the inner heliosphere.

In considering the growth rate of the PDI relative to the linear wave time, $\omega_0,$ we do not take into account the effect of turbulence solar wind. The non-linear interaction time between Alfv\'enic fluctuations likely plays an important role in parametric coupling. Specifically, it is likely that a critically balanced cascade with non-linear interactions occuring on order the wave propagation time could disrupt parametric coupling between Alfv\'enic and slow-mode like fluctuations \cite{Schekochihin2009}. This idea has been touched upon in \citet{Shietal2017}, in which reduced PDI growth rates were found in a turbulent plasma. Our future work looks to address the physics of parametric coupling in a turbulent plasma and to compare the growth of compressive modes through PDI versus their recycling through a turbulent cascade.

\acknowledgments
The authors would like to acknowledge helpful discussions with Marco Velli, Anna Tenerani, and Alfred Mallet. We would additionally like to extend our thanks to Marc Pulupa and Chadi Salem for providing processed electron data from Wind/3DP. T.A.B. is supported by NASA Earth and Space Science Fellowship NNX16AT22H. P.H. acknowledges grant 18-08861S of the Czech Science Foundation. Wind/3DP data analysis at UC Berkeley is supported in part by NASA grant NNX16AP95G.

\begin{thebibliography}{}
\expandafter\ifx\csname natexlab\endcsname\relax\def\natexlab#1{#1}\fi

\bibitem[{{Araneda} {et~al.}(2009){Araneda}, {Maneva}, \&
  {Marsch}}]{Aranedaetal2009}
{Araneda}, J.~A., {Maneva}, Y., \& {Marsch}, E. 2009, Physical Review Letters,
  102, 175001

\bibitem[{Araneda {et~al.}(2008)Araneda, Marsch, \&
  F.-Vi\~nas}]{Aranedaetal2008}
Araneda, J.~A., Marsch, E., \& F.-Vi\~nas, A. 2008, Phys. Rev. Lett., 100,
  125003

\bibitem[{{Barnes}(1966)}]{Barnes1966}
{Barnes}, A. 1966, Physics of Fluids, 9, 1483

\bibitem[{{Chandran}(2017)}]{Chandran2017}
{Chandran}, B.~D.~G. 2017, ArXiv e-prints, arXiv:1712.09357

\bibitem[{{Chen} {et~al.}(2012){Chen}, {Mallet}, {Schekochihin}, {Horbury},
  {Wicks}, \& {Bale}}]{Chen3danis2012}
{Chen}, C.~H.~K., {Mallet}, A., {Schekochihin}, A.~A., {et~al.} 2012, \apj,
  758, 120

\bibitem[{{Del Zanna}(2001)}]{DelZanna2001}
{Del Zanna}, L. 2001, \grl, 28, 2585

\bibitem[{{Del Zanna} {et~al.}(2015){Del Zanna}, {Matteini}, {Landi},
  {Verdini}, \& {Velli}}]{DelZannaetal2015}
{Del Zanna}, L., {Matteini}, L., {Landi}, S., {Verdini}, A., \& {Velli}, M.
  2015, Journal of Plasma Physics, 81, 325810102

\bibitem[{{Del Zanna} {et~al.}(2001){Del Zanna}, {Velli}, \&
  {Londrillo}}]{DelZannaetal2001}
{Del Zanna}, L., {Velli}, M., \& {Londrillo}, P. 2001, \aap, 367, 705

\bibitem[{{Derby}(1978)}]{Derby1978}
{Derby}, Jr., N.~F. 1978, \apj, 224, 1013

\bibitem[{{Goldstein}(1978)}]{Goldstein1978}
{Goldstein}, M.~L. 1978, \apj, 219, 700

\bibitem[{{Hollweg}(1994)}]{Hollweg1994}
{Hollweg}, J.~V. 1994, \jgr, 99, 23

\bibitem[{{Horbury} {et~al.}(2008){Horbury}, {Forman}, \&
  {Oughton}}]{Horburyetal2008}
{Horbury}, T.~S., {Forman}, M., \& {Oughton}, S. 2008, Physical Review Letters,
  101, 175005

\bibitem[{{Howes} {et~al.}(2012){Howes}, {Bale}, {Klein}, {Chen}, {Salem}, \&
  {TenBarge}}]{Howesetal2012}
{Howes}, G.~G., {Bale}, S.~D., {Klein}, K.~G., {et~al.} 2012, \apjl, 753, L19

\bibitem[{{Jayanti} \& {Hollweg}(1993)}]{JayantiHollweg1993}
{Jayanti}, V., \& {Hollweg}, J.~V. 1993, \jgr, 98, 19

\bibitem[{{Kauffmann} \& {Araneda}(2008)}]{KauffmanAraneda2008}
{Kauffmann}, K., \& {Araneda}, J.~A. 2008, Physics of Plasmas, 15, 062106

\bibitem[{Kellogg \& Horbury(2005)}]{KelloggHorbury2005}
Kellogg, P.~J., \& Horbury, T.~S. 2005, Annales Geophysicae, 23, 3765

\bibitem[{{Klein} {et~al.}(2012){Klein}, {Howes}, {TenBarge}, {Bale}, {Chen},
  \& {Salem}}]{Kleinetal2012}
{Klein}, K.~G., {Howes}, G.~G., {TenBarge}, J.~M., {et~al.} 2012, \apj, 755,
  159

\bibitem[{{Lepping} {et~al.}(1995){Lepping}, {Ac{\~u}na}, {Burlaga}, {Farrell},
  {Slavin}, {Schatten}, {Mariani}, {Ness}, {Neubauer}, {Whang}, {Byrnes},
  {Kennon}, {Panetta}, {Scheifele}, \& {Worley}}]{LeppingMFI}
{Lepping}, R.~P., {Ac{\~u}na}, M.~H., {Burlaga}, L.~F., {et~al.} 1995, \ssr,
  71, 207

\bibitem[{{Lin} {et~al.}(1995){Lin}, {Anderson}, {Ashford}, {Carlson},
  {Curtis}, {Ergun}, {Larson}, {McFadden}, {McCarthy}, {Parks}, {R{\`e}me},
  {Bosqued}, {Coutelier}, {Cotin}, {D'Uston}, {Wenzel}, {Sanderson}, {Henrion},
  {Ronnet}, \& {Paschmann}}]{Lin3DP}
{Lin}, R.~P., {Anderson}, K.~A., {Ashford}, S., {et~al.} 1995, \ssr, 71, 125

\bibitem[{{Malara} {et~al.}(2001){Malara}, {Primavera}, \&
  {Veltri}}]{Malaraetal2001}
{Malara}, F., {Primavera}, L., \& {Veltri}, P. 2001, Nonlinear Processes in
  Geophysics, 8, 159

\bibitem[{{Matteini} {et~al.}(2010){Matteini}, {Landi}, {Del Zanna}, {Velli},
  \& {Hellinger}}]{Matteinietal2010}
{Matteini}, L., {Landi}, S., {Del Zanna}, L., {Velli}, M., \& {Hellinger}, P.
  2010, \grl, 37, L20101

\bibitem[{{Narita} \& {Marsch}(2015)}]{NaritaMarsch2015}
{Narita}, Y., \& {Marsch}, E. 2015, \apj, 805, 24

\bibitem[{{Ogilvie} {et~al.}(1995){Ogilvie}, {Chornay}, {Fritzenreiter},
  {Hunsaker}, {Keller}, {Lobell}, {Miller}, {Scudder}, {Sittler}, {Torbert},
  {Bodet}, {Needell}, {Lazarus}, {Steinberg}, {Tappan}, {Mavretic}, \&
  {Gergin}}]{OgilvieSWE}
{Ogilvie}, K.~W., {Chornay}, D.~J., {Fritzenreiter}, R.~J., {et~al.} 1995,
  \ssr, 71, 55

\bibitem[{Pulupa {et~al.}(2014)Pulupa, Bale, Salem, \&
  Horaites}]{Pulupaetal2014}
Pulupa, M.~P., Bale, S.~D., Salem, C., \& Horaites, K. 2014, Journal of
  Geophysical Research: Space Physics, 119, 647

\bibitem[{{Schekochihin} {et~al.}(2009){Schekochihin}, {Cowley}, {Dorland},
  {Hammett}, {Howes}, {Quataert}, \& {Tatsuno}}]{Schekochihin2009}
{Schekochihin}, A.~A., {Cowley}, S.~C., {Dorland}, W., {et~al.} 2009, \apjs,
  182, 310

\bibitem[{{Shi} {et~al.}(2017){Shi}, {Li}, {Xiao}, \& {Wang}}]{Shietal2017}
{Shi}, M., {Li}, H., {Xiao}, C., \& {Wang}, X. 2017, \apj, 842, 63

\bibitem[{{Tenerani} \& {Velli}(2013)}]{TeneraniVelli2013}
{Tenerani}, A., \& {Velli}, M. 2013, Journal of Geophysical Research (Space
  Physics), 118, 7507

\bibitem[{{Tenerani} {et~al.}(2017){Tenerani}, {Velli}, \&
  {Hellinger}}]{Teneranietal2017}
{Tenerani}, A., {Velli}, M., \& {Hellinger}, P. 2017, \apj, 851, 99

\bibitem[{Tu \& Marsch(1994)}]{TuMarsch1994}
Tu, C.~Y., \& Marsch, E. 1994, Journal of Geophysical Research: Space Physics,
  99, 21481

\bibitem[{{Verscharen} {et~al.}(2017){Verscharen}, {Chen}, \&
  {Wicks}}]{Verscharenetal2017}
{Verscharen}, D., {Chen}, C.~H.~K., \& {Wicks}, R.~T. 2017, \apj, 840, 106

\bibitem[{{Yao} {et~al.}(2011){Yao}, {He}, {Marsch}, {Tu}, {Pedersen},
  {R{\`e}me}, \& {Trotignon}}]{Yaoetal2011}
{Yao}, S., {He}, J.-S., {Marsch}, E., {et~al.} 2011, \apj, 728, 146

\bibitem[{{Yao} {et~al.}(2013{\natexlab{a}}){Yao}, {He}, {Tu}, {Wang}, \&
  {Marsch}}]{Yaoetal2013a}
{Yao}, S., {He}, J.-S., {Tu}, C.-Y., {Wang}, L.-H., \& {Marsch}, E.
  2013{\natexlab{a}}, \apj, 776, 94

\bibitem[{{Yao} {et~al.}(2013{\natexlab{b}}){Yao}, {He}, {Tu}, {Wang}, \&
  {Marsch}}]{Yaoetal2013b}
---. 2013{\natexlab{b}}, \apj, 774, 59

\end{thebibliography}
\end{document}